\documentclass[referee]{raa}           

\usepackage{graphicx,times}             
\usepackage{natbib}
\usepackage{amssymb,amsmath}
\bibpunct{(}{)}{;}{a}{}{,}

\usepackage{hyperref}
\hypersetup{colorlinks = true, linkcolor = green, anchorcolor = red, citecolor = blue, filecolor = red, pagecolor = red, urlcolor = red}

\begin{document}

   \title{Precision of The Chinese Space Station Telescope (CSST) Stellar Radial Velocities
}

   \volnopage{Vol.0 (20xx) No.0, 000--000}      
   \setcounter{page}{1}          

   \author{Yang Sun
      \inst{1}
   \and Ding-Shan Deng
      \inst{1}
   \and Hai-Bo Yuan
      \inst{1}
   }

   \institute{Department of Astronomy,
   Beijing Normal University, Beijing 100875, China; {\it yuanhb@bnu.edu.cn}\\
\vs\no
   {\small Received~~20xx month day; accepted~~20xx~~month day}}

\abstract{The Chinese Space Station Telescope (CSST) spectroscopic survey plans to deliver high-quality low-resolution ($R > 200$) slitless spectra
for hundreds of millions of targets down to a limiting magnitude of about 21 mag, covering a large survey area (17500 deg$^2$) and
a wide wavelength range (255-1000 nm by 3 bands GU, GV, and GI).
In this work,  we use  empirical spectra of the Next Generation Spectral Library to simulate the CSST stellar spectra at $R = 250$,
and investigate their capabilities in measuring radial velocities.
We find that velocity uncertainties depend strongly on effective temperature, weakly on metallicity for only FGK stars, and
hardly on surface gravity. It is possible to deliver stellar radial velocities to a precision of about $3 \,\mathrm{km}\,\mathrm{s}^{-1}$ for
AFGKM stars, and about $10 \,\mathrm{km}\,\mathrm{s}^{-1}$ for OB stars, at signal-to-noise ratio (SNR) of 100.
Velocity uncertainties using single GU/GV/GI band spectra are
also explored. Given the same SNR, the GU band performs best, the GV band the second best, and then the GI band.
The effects of spectral normalization and imperfect template on velocity measurements are investigated and found to be very weak.
The uncertainties caused by wavelength calibration are considered and found to be moderate. 
Given the possible precision of radial velocities, the CSST spectroscopic survey can enable interesting science such as searching for hyper-velocity stars.
Limitations of our results are also discussed.
\keywords{methods: data analysis — methods: statistical — stars: fundamental parameters — stars: kinematics and dynamics}
}

   \authorrunning{Y. Sun, D.-S. Deng \& H.-B. Yuan }            
   \titlerunning{Precision of The CSST Stellar Radial Velocities }  

   \maketitle

%
%
\section{Introduction}           
\label{sect:intro}

The Chinese Space Station Telescope (CSST)  Optical Survey (CSS-OS; \citealt{zhan_consideration_2011, 2018cosp...42E3821Z}; \citealt{cao_testing_2018}; \citealt{gong_cosmology_2019})
is one of the leading astronomical projects in China. With the CSST, a 2 meter space telescope of a large field of view of 1.1 ${\mathrm{deg}^{2}}$   and sharing the
same orbit of the Chinese Space Station, it will simultaneously carry out both  photometric and slitless grating spectroscopic surveys, covering a large sky area of
17500 ${\mathrm{deg}^{2}}$ but at a high spatial resolution of $\sim 0.15''$ in about ten years.
The imaging survey has 7 photometric filters, i.e. $NUV , u, g, r, i, z$, and $y$ bands, covering 255 --1000 nm from the near-ultraviolet (NUV) to near infrared (NIR).
While the slitless spectroscopic survey has 3 bands, GU (255-420 nm), GV (400-650 nm), and GI (620-1000 nm), at a spectral resolving power  larger than 200.
The main scientific goals of the CCS-OS include the nature of dark matter and dark energy, large scale structure and cosmology,  galaxy formation and evolution,
active galactic nucleus (AGNs), the Milky Way and near-field cosmology, stellar physics, solar system, and astrometry.

Given the very low sky background and very high spatial resolution,
space-based wide-field slitless spectroscopic surveys  are very powerful tools in astronomy \citep{glazebrook_monster_2005}.
Complementary to the CSST imaging survey, the CSST spectroscopic survey aims to deliver high quality slitless spectra covering 250 -- 1000 nm at R  larger than 200
for hundreds of millions of stars and galaxies. Such a large and magnitude-limited sample of stars will provide  a great opportunity for studies of stars and the Galaxy.
However, precise stellar parameters have to be extracted from the unique CSST low resolution spectra first. It is shown that \citep{ting_prospects_2017} stellar parameters (e.g., effective temperatures, surface gravities and elemental abundances) can be determined almost to the same precision for different spectral resolutions (R $>$  1000),
given perfect spectral models, spectral normalization,  the same exposure time and number of detector pixels.
At lower resolutions, estimates of some stellar parameters may be strongly correlated with one another.
Therefore, it is very essential to explore capabilities of  the  unique CSS-OS slitless spectra in deriving stellar parameters.
Considering the relative simplicity in measuring radial velocities, we focus on stellar radial velocities in this paper and leave other parameters in future work.

In this work, we estimate precision of radial velocity measurements from CSS-OS stellar spectra.
Spectra from the Next Generation Spectral Library (NGSL; \citealt{2007ASPC..374..409H}) are used to simulate the CSS-OS spectra.
A cross-correlation function (CCF) method \citep{1996A&AS..119..373B} is used to measure radial velocities, and Monte Carlo simulations are used to compute their errors.
Variations of precision of radial velocities as functions of stars' atmospheric parameters are also investigated.
The paper is organized as follows: We introduce our data and method in Section 2 and 3, respectively. The results are shown in Section 4.
In Section 5, we discuss implications and limitations of our results. We conclude in Section 6.

\section{Data}
\label{sect:data}

In this work, we use spectra from the  NGSL (\citealt{2007ASPC..374..409H}; \citealt{koleva_stellar_2012}) to simulate the CSS-OS slitless spectroscopic observations.
The NGSL is a empirical stellar spectral library constructed by a HST/STIS snapshot program with
the three gratings G230LB+G430L+G750L.
It contains intermediate resolution ($R \sim 1000$), UV-optical spectra (1670-10250 \AA) of 378 stars spanning a wide range in temperature, luminosity, and metallicity.
The spectra in this library are well calibrated and have typical SNRs of 200, 600, and 400 per pixel for the GU, GV and GI bands, respectively.
Due to the  good match in wavelength coverage and wide coverage in stellar types,
the NGSL provides  an ideal empirical dataset for the purpose of this work.

\begin{figure}[!h]
\centering
\includegraphics[width=\textwidth, angle=0]{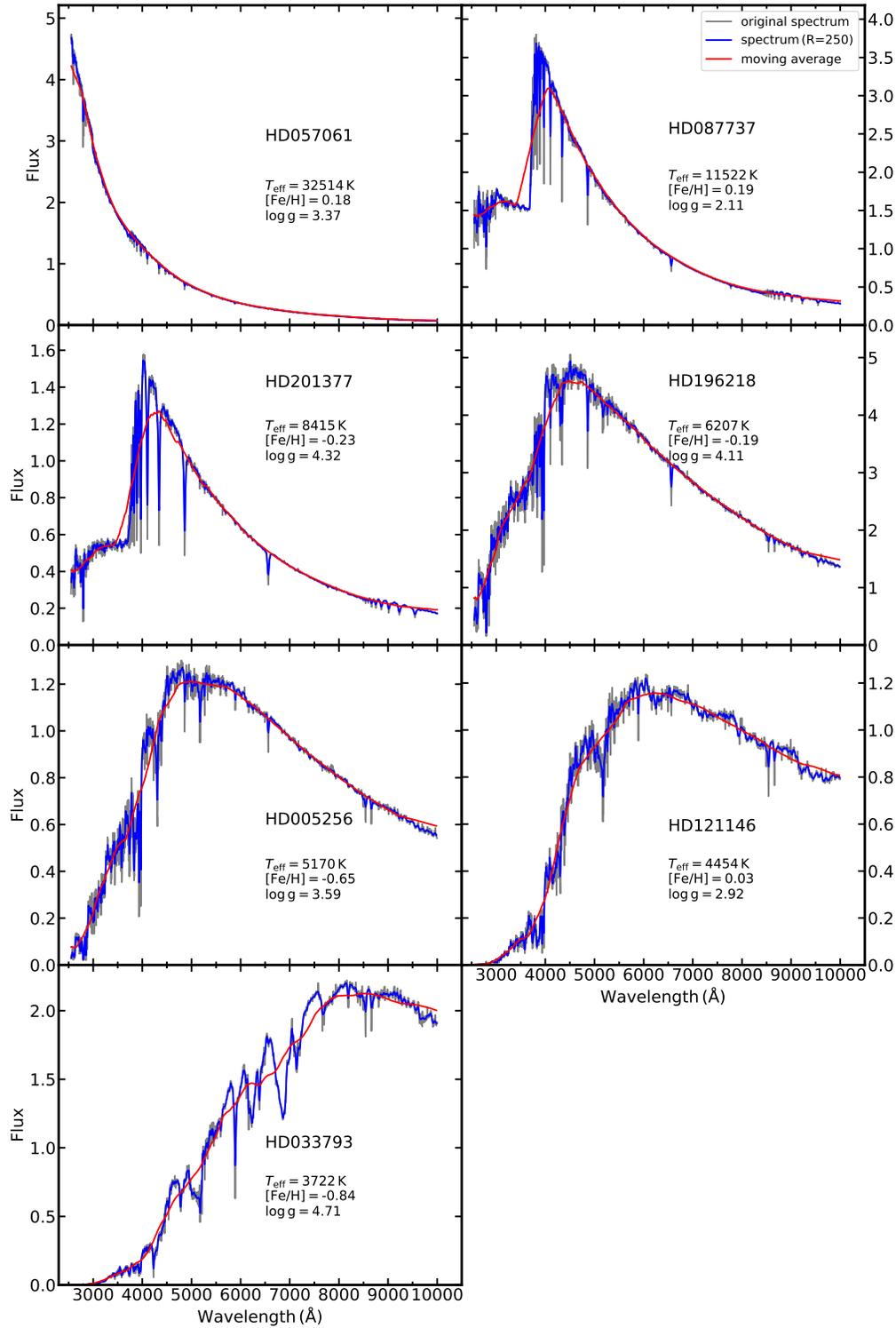}
\caption{NGSL spectra of representative stars of different spectral types.  The Y-axes are in arbitrary units. Stellar ID and parameters are marked for each target.
The grey and blue lines denote the original and degraded  ($R=250$) spectra, respectively.
The red lines are continua of the degraded spectra.}
\label{Fig1}
\end{figure}

\begin{figure}[!h]
\centering
\includegraphics[width=\textwidth, angle=0]{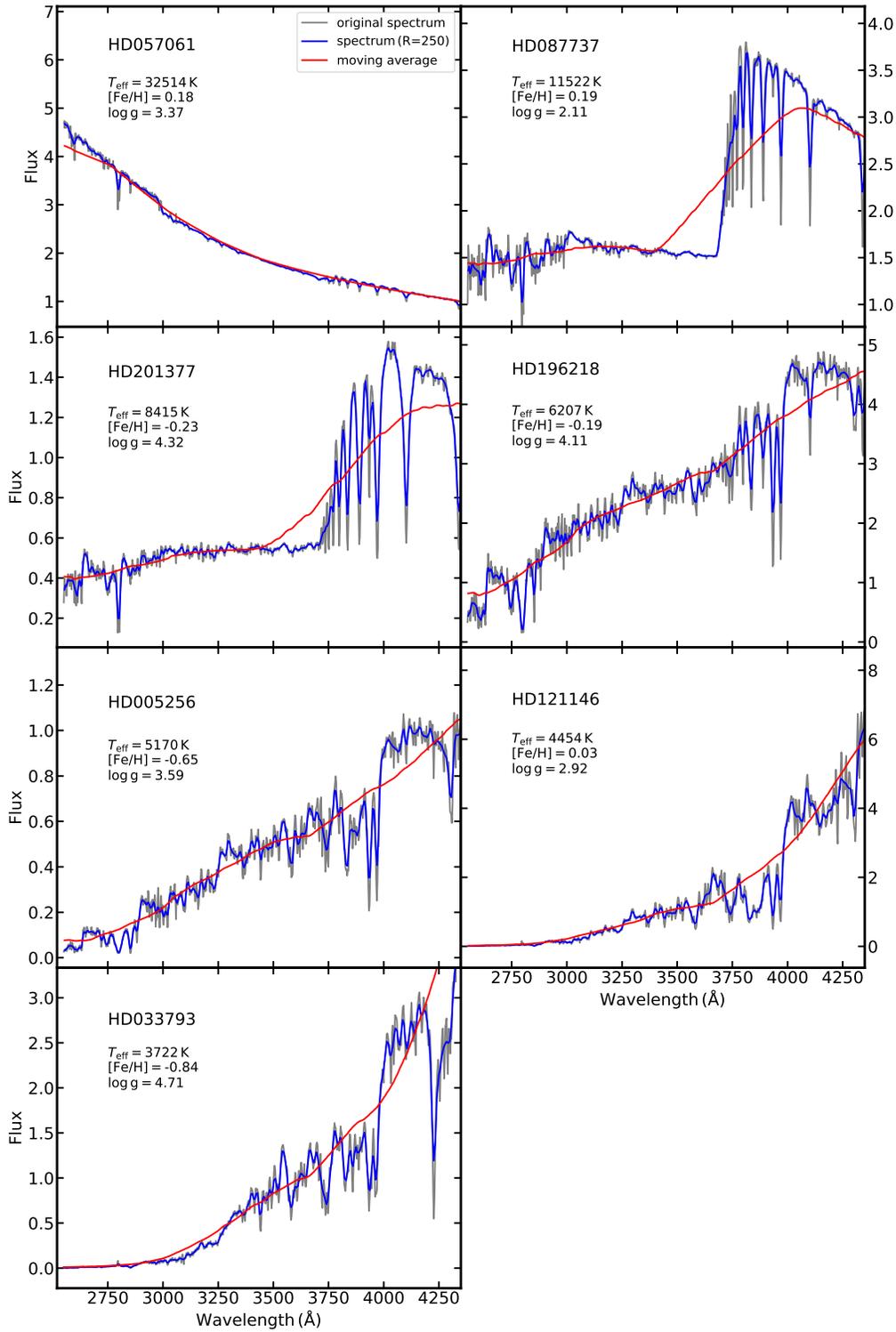}
\caption{Same to Fig.\,1 but for the GU band spectra only.}
\label{Fig2}
\end{figure}

The NGSL spectra have a higher spectral resolution than the CSS-OS slitless spectra ($R > 200$). They are degraded to $R = 250$ using a Gaussian Convolution method.
Here we adopt a constant spectral resolution of $R=250$ and a sampling rate of 3 pixels per resolution element for the CSS-OS spectra. Fig.~\ref{Fig1} and Fig.~\ref{Fig2} show the comparison between the original ($R \sim 1000$)  and degraded ($R = 250$)  NGSL spectra for stars of different spectral types.
Strong or broad stellar absorption features, such as Balmer Jump (BJ), Balmer lines, Paschen lines, the D4000 break, Mg II doublet at 2800\AA,  Mg I at 2852\AA, the Mg break at $\sim$2600\AA, the CH G bands at 4300\AA, MgH at 5150\AA, Na I D1 and D2 lines, Ca II H$\&$K lines, Ca II IR triplets and TiO bands, are still clearly visible in the degraded spectra.
Weak and narrow absorption lines are mixed and hardly seen.

\section{Method}
\label{sect:method}

\begin{figure}[!h]
\centering
\includegraphics[width=100mm,height=70mm, angle=0]{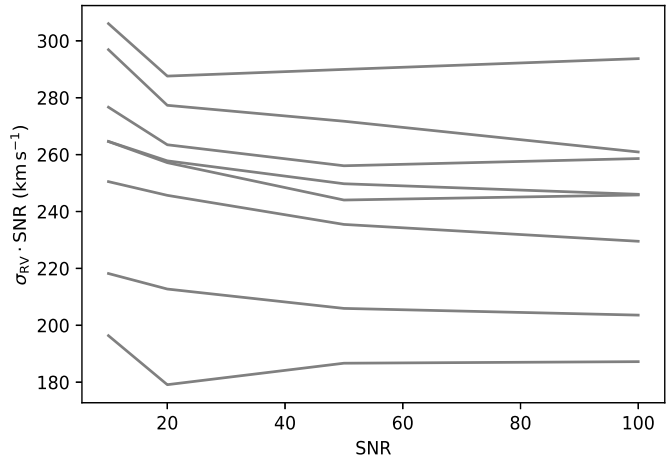}
\caption{The inverse relationship between $\sigma_{\mathrm{RV}}$ and SNR for 8 randomly selected stars of different spectral types.}
\label{Fig3}
\end{figure}

We use the cross-correlation function (CCF) method to measure stellar radial velocities and Monte Carlo simulations to estimate their errors.
We assume that the degraded NGSL spectra have zero noises and are used as templates.
For each target from the NGSL, 1,000 simulated spectra for a given SNR are obtained by adding random noises to each pixel of its degraded spectrum.
We then measure velocities by a Gaussian fit to the CCF of the simulated spectra with its template spectrum \citep{1996A&AS..119..373B}.
Dispersion value of its 1000 velocity measurements is then calculated to represent its velocity precision ($\sigma_{\mathrm{RV}}$) at the given SNR.
For a given spectrum, $\sigma_{\mathrm{RV}}$ is inversely proportional to its SNR  (Fig.~\ref{Fig3}).
Therefore, only one single SNR of 100 is adopted in the work.  Note here the SNR refers to SNR per pixel at a sampling rate of 3 pixels per resolution element.

In the above procedure,  we have assumed  that all CSS-OS slitless spectra are accurately  flux calibrated and reddening corrected,
only then simulated spectra can be directly compared to  temples without normalization.
We believe that we can achieve $\sim$ 1 per cent flux calibration for the CSS-OS slitless spectroscopic survey, with an innovative technique
that uses tens of millions of stars as flux standards (\citealt{Yuan_CSSTwave_tobesubmitted}).
Reddening correction is relatively easy due to the fact that the CSS-OS will mostly target high Galactic latitude regions by using the popular two-dimensional dust extinction maps (\citealt{schlegel_application_1998}, \citealt{planck_collaboration_planck_2014}; \citealt{irfan_determining_2019}; Sun et al. in prep).
In cases of bad flux calibration or difficult reddening correction, observed spectra have to be normalized first to measure their velocities.
Therefore, we also explore velocity precision using normalized CSS-OS spectra in the work. 

However, it is non-trivial to do spectral normalization due to difficulties in defining clean continuum areas, particularly in the cases of very low spectral resolution.
We adopt a simple moving average method to estimate the continuum of each spectrum, with a window size of 51 pixels.
In cases of boundary pixels,  where there are less than 25 pixels in one side, all pixels in that side are used.
Despite its simplicity, this approach is robust, easy to be implemented and retains most features.
Fig.~\ref{Fig1} and Fig.~\ref{Fig2} show the obtained continua for representative stars of different spectral types.

The CSS-OS slitless spectrograph has 3 filters, GU (255-420 nm), GV (400-650 nm) , and GI (620-1000 nm).
For a given target, the 3 filters will be observed at different epochs. 
It means that only one segment of spectrum (GU/GV/GI) will be available in the early stage of the survey.
Therefore, we also explore velocity precision using only one segment of CSS-OS spectra in the work.

\section{Result}
\label{sect:R}

\subsection{Based on the whole spectra}

\begin{figure}[!h]
\centering
\includegraphics[width=140mm, angle=0]{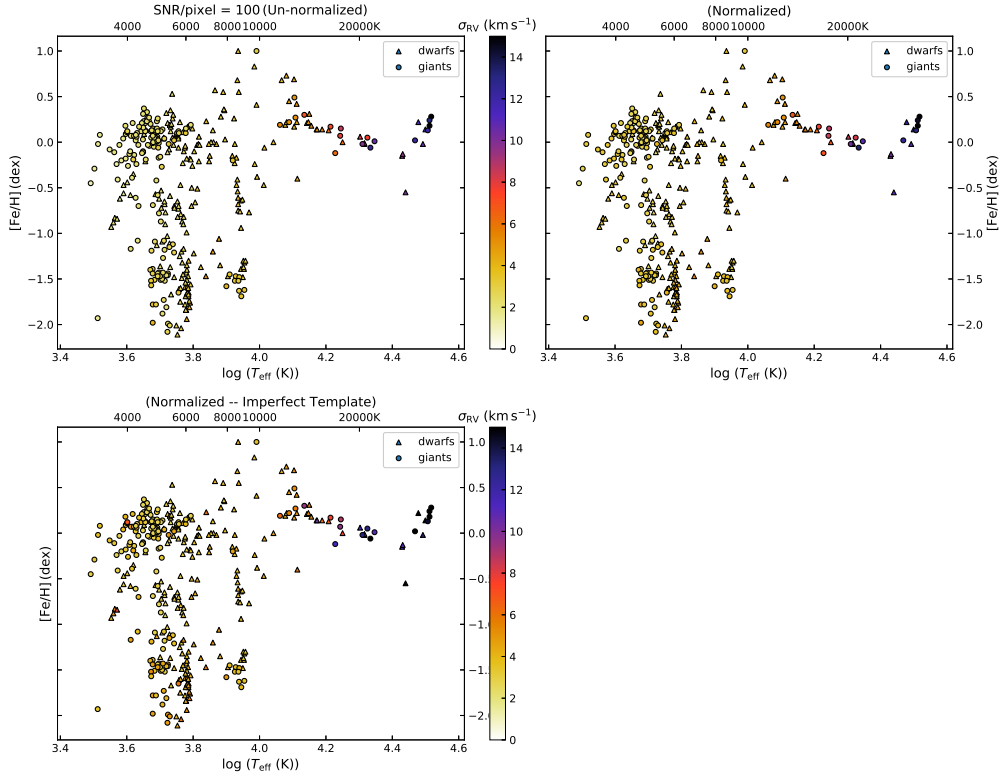}
\caption{The relationship between $\sigma_{\mathrm{RV}}$ and stellar parameters at SNR=100 
for the un-normalized spectra (upper-left), 
the normalized spectra (upper-right), 
and the normalized spectra using  the closest spectrum as template (lower-left).
The triangles and circles denote dwarfs and giants, respectively. The color bar represents the values of $\sigma_{\mathrm{RV}}$.}
\label{Fig4}
\end{figure}

\begin{figure}[!h]
   \centering
   \includegraphics[width=140mm, angle=0]{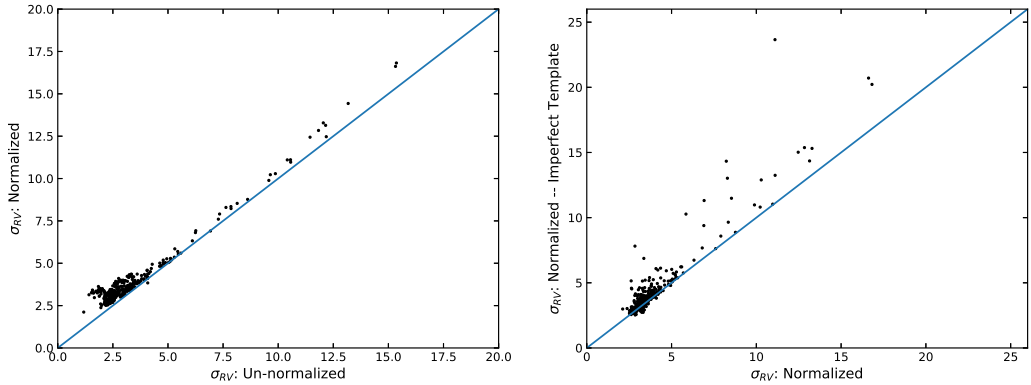}
   \caption{Comparisons of $\sigma_{\mathrm{RV}}$ results obtained by different methods.
   The blue solid line in each panel represents the line of equality.}
   \label{Fig5}
   \end{figure}

We first investigate velocity uncertainties ($\sigma_{\mathrm{RV}}$) with the full range of spectra at SNR=100.
The upper-left panel of Fig.~\ref{Fig4} plots the relationship between $\sigma_{\mathrm{RV}}$ and stellar parameters ($T_{\mathrm{eff}}$, $\mathrm{[Fe/H]}$ and $\log \mathrm{g}$).
The triangles and circles denote dwarfs ($\log \mathrm{g} \ge 3.5$ dex) and giants ($\log \mathrm{g} < 3.5$ dex), respectively.
It can be seen that $\sigma_{\mathrm{RV}}$ depends strongly on spectral types.
The $\sigma_{\mathrm{RV}}$ values are about
$2 - 4 \,\mathrm{km}\,\mathrm{s}^{-1}$ for AFGKM  stars, increase to $ 4 -  15 \,\mathrm{km}\,\mathrm{s}^{-1}$ for hot OB stars.
There is a weak anti-correlation between $\mathrm{[Fe/H]}$ and $\sigma_{\mathrm{RV}}$ for FGK stars, as more metal-rich
stars display stronger metal absorption lines. While $\log \mathrm{g}$ has no effects on $\sigma_{\mathrm{RV}}$.
The upper-right panel of Fig.~\ref{Fig4} shows the result from normalized spectra .
The trends are very similar, but with slightly larger $\sigma_{\mathrm{RV}}$ values due to loss of information in the normalization process.
The upper panels of Fig.~\ref{Fig5} compare the results between unnormalized spectra and normalized spectra.  
The uncertainties are only slightly larger with normalized spectra, suggesting that the effects of spectral normalization on velocity measurements are weak.

\subsection{Based on GU/GV/GI band spectra}

\begin{figure}[!h]
\centering
\includegraphics[width=140mm, angle=0]{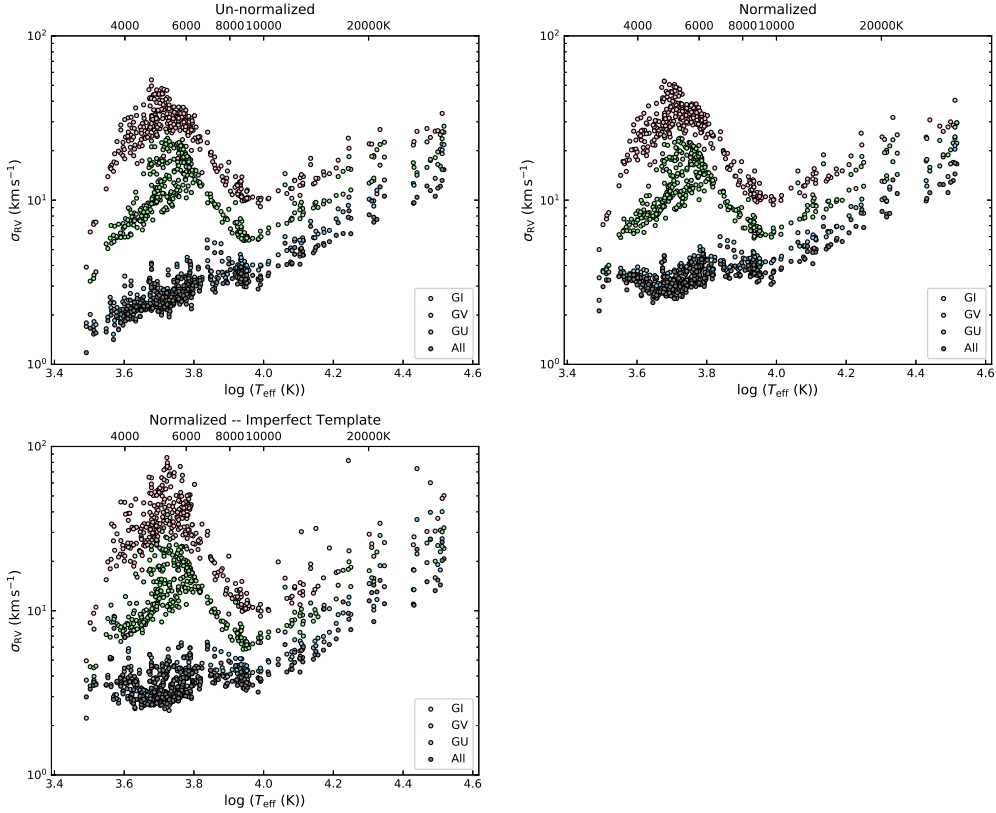}
\caption{The relationship between  $\sigma_{\mathrm{RV}}$  and $T_{\mathrm{eff}}$ at SNR=100 for the un-normalized spectra (upper-left), the normalized spectra (upper-right), and the normalized spectra and the closest spectrum as template (lower-left).
The red, green, blue and black dots represent results measured using the GI, GV, GU, and the full spectra, respectively.}
\label{Fig6}
\end{figure}

We also investigate velocity uncertainties using each segment of CSS-OS spectra. 
Fig.~\ref{Fig6} plots $\sigma_{\mathrm{RV}}$  as a function of  $T_{\mathrm{eff}}$, using spectra from the GI (red dots), GV (green dots), GU (blue dots) bands and the whole spectra (black dots).
Similarly, the upper-left, upper-right, and lower-left  panels show results from the un-normalized spectra,  the normalized spectra, and 
the normalized spectra using the closest spectrum as template, respectively.
We divide the sample into different $T_{\mathrm{eff}}$ bins. The bin size is 500 K for $T_{\mathrm{eff}}$ from 3000 to 10000 K, and is 2000 K for hotter stars.
The median values of $\sigma_{\mathrm{RV}}$
are estimated for each bin and listed in Table~\ref{Tab:sigmaRV-Teff}.

The upper-left panel of Fig.~\ref{Fig6} shows that the GU band plays a key role in measuring stellar radial velocities. At the same SNR,
$\sigma_{\mathrm{RV (GU)}} <  \sigma_{\mathrm{RV (GV)}}  < \sigma_{\mathrm{RV (GI)}} $  for all types of stars, due to more absorption features in the blue.
But note in real cases, due to the usually lower SNRs of the GU band spectra than those of the GV and GI bands, the precision of velocity measurements would not be determined by the GU band. All the three bands shall play different roles in measuring velocities.
$\sigma_{\mathrm{RV}}$ depends strongly on $T_{\mathrm{eff}}$. 
For the GI band, $\sigma_{\mathrm{RV (GI)}}$  is lowest, about $ 7  \,\mathrm{km}\,\mathrm{s}^{-1}$, for very cool stars due to strong molecular absorption bands in the spectra.
Then the values increase rapidly to about  $ 30  \,\mathrm{km}\,\mathrm{s}^{-1}$ at $T_{\mathrm{eff}}$ around 4,500 K, then decrease slowly to about $ 10  \,\mathrm{km}\,\mathrm{s}^{-1}$ at $T_{\mathrm{eff}}$ around 10,000 K due to stronger Balmer lines and increase again to about $ 25 \,\mathrm{km}\,\mathrm{s}^{-1}$ at $T_{\mathrm{eff}}$ around 33,000 K.
The large scattering around FGK stars is caused by different metallicities.
For the GV band, it shows the similar trend with the GI band. It performs best both for cool M stars (about $ 4  \,\mathrm{km}\,\mathrm{s}^{-1}$  ) due to strong molecular absorption bands
and for A stars (about $ 6  \,\mathrm{km}\,\mathrm{s}^{-1}$ ) due to strong Balmer lines.
Thanks to different strong features for different types of stars in the GU band, such as Balmer lines and BJ for hot stars, Mg II doublet at 2800\AA, Ca H$\&$K lines and D4000 break for cool stars, $\sigma_{\mathrm{RV (GU)}}$
increases almost monotonically with $T_{\mathrm{eff}}$, from  $ 2 \,\mathrm{km}\,\mathrm{s}^{-1}$  for M stars to $ 20  \,\mathrm{km}\,\mathrm{s}^{-1}$ for O stars.
The upper-right and lower-left panels in Fig.~\ref{Fig6} show similar trends.

Note that eight stars, $\mathrm{HD 190073}$, $\mathrm{GL 109}$, $\mathrm{LHS 10}$, $\mathrm{LHS 482}$, G 260-36, CD-691618, $\mathrm{HD 188262}$ and $\mathrm{HD 193495}$, are excluded in the above analysis.
For the first six sources, their spectra in the GU band are noisy due to  low SNRs.
For $\mathrm{HD 193495}$ and $\mathrm{HD 188262}$, their GU band spectra show much higher fluxes at 2550-3000 \AA$\,$ than expected.

\begin{table}[!h]
\bc

\caption[]{Predicted uncertainties of the CSST stellar radial velocities as a function of effective temperature at SNR = 100}
\label{Tab:sigmaRV-Teff}

\setlength{\tabcolsep}{3pt}
\small
\begin{tabular}{ccccccccccccccccc}
   \hline
   $T_{\mathrm{eff}}$ (K) & \multicolumn{16}{c}{$\sigma_{\mathrm{RV}}$ ($\mathrm{km}\,\mathrm{s}^{-1}$)}                                                                                                                                \\ \cline{2-17}
         & \multicolumn{4}{c}{un-normalized}&  &  & \multicolumn{4}{c}{normalized} &  &  & \multicolumn{4}{c}{normalized$^a$} \\
         \cline{2-5} \cline{8-11} \cline{14-17} 
         & GU     & GV     & GI     & All    &  &  & GU       & GV       & GI       & All     &  &  & GU       & GV       & GI       & All     \\ \hline
   3250  & 1.8    & 3.3    & 7.1    & 1.6    &   &  & 3.7      & 3.7      & 7.7      & 3.3     &  &  & 3.7      & 4.5      & 8.4      & 3.5     \\
   3750  & 2.1    & 6.1    & 18.9   & 1.9    &   &  & 3.5      & 6.9      & 18.6     & 3.3     &  &  & 3.6      & 7.9      & 22.7     & 3.6     \\
   4250  & 2.2    & 7.6    & 24.7   & 2.2    &  &  & 3.1      & 8.2      & 25.4     & 3.0     &  &  & 3.3      & 8.3      & 26.5     & 3.1     \\
   4750  & 2.4    & 10.2   & 30.3   & 2.3    &  &  & 3.0      & 10.4     & 30.6     & 2.8     &  &  & 3.1      & 10.9     & 33.3     & 2.9     \\
   5250  & 2.5    & 13.0   & 35.0   & 2.5    &  &  & 3.2      & 13.9     & 35.4     & 3.0     &  &  & 3.2      & 14.9     & 45.0     & 3.1     \\
   5750  & 2.7    & 16.4   & 31.4   & 2.6    &   &  & 3.4      & 16.5     & 32.5     & 3.2     &  &  & 3.4      & 18.8     & 39.6     & 3.3     \\
   6250  & 3.1    & 15.8   & 28.0   & 3.0    &   &  & 4.0      & 15.9     & 30.1     & 4.0     &  &  & 4.1      & 17.0     & 38.1     & 4.0     \\
   6750  & 3.3    & 12.3   & 23.0   & 3.2    &  &  & 3.8      & 12.5     & 24.5     & 3.6     &  &  & 3.9      & 13.1     & 27.3     & 3.8     \\
   7250  & 3.5    & 9.5    & 17.1   & 3.3    &   &  & 4.0      & 9.7      & 17.9     & 3.7     &  &  & 4.3      & 9.8      & 19.6     & 4.0     \\
   7750  & 3.6    & 7.7    & 13.8   & 3.2    & &  & 4.0      & 8.5      & 14.0     & 3.6     &  &  & 4.2      & 8.3      & 15.9     & 3.8     \\
   8250  & 3.9    & 6.7    & 11.3   & 3.4    &  &  & 4.2      & 7.0      & 12.0     & 3.7     &  &  & 4.4      & 7.0      & 12.1     & 3.8     \\
   8750  & 3.9    & 6.0    & 10.5   & 3.3    &  &  & 4.2      & 6.5      & 10.8     & 3.6     &  &  & 4.3      & 6.5      & 10.8     & 3.7     \\
   9250  & 3.8    & 5.8    & 10.3   & 3.2    &   &  & 4.2      & 6.1      & 10.4     & 3.5     &  &  & 4.1      & 6.2      & 10.6     & 3.6     \\
   9750  & 3.8    & 6.0    & 10.0   & 3.3    &   &  & 4.2      & 6.3      & 10.0     & 3.4     &  &  & 4.3      & 6.8      & 10.1     & 3.8     \\
   11000 & 4.8    & 7.2    & 11.2   & 4.0    &  &  & 5.2      & 7.5      & 11.5     & 4.2     &  &  & 5.1      & 7.8      & 11.9     & 4.3     \\
   13000 & 5.7    & 8.7    & 12.7   & 4.7    &  &  & 5.9      & 9.2      & 13.3     & 5.0     &  &  & 5.9      & 8.8      & 13.1     & 5.1     \\
   15000 & 6.4    & 9.6    & 13.9   & 5.3    &  &  & 6.6      & 9.8      & 14.2     & 5.5     &  &  & 6.7      & 10.2     & 20.0     & 5.7     \\
   17000 & 8.7    & 14.4   & 16.6   & 6.8    &  &  & 9.1      & 14.8     & 17.8     & 7.3     &  &  & 9.6      & 16.0     & 20.8     & 8.3     \\
   19000$^b$ &        &        &        &        &  &  &        &       &       &       &  &  &          &          &          &           \\
   21000 & 10.8   & 16.0   & 20.3   & 8.0    &   &  & 11.0     & 16.8     & 21.8     & 8.4     &  &  & 13.2     & 20.2     & 31.7     & 13.0    \\
   23000 & 13.3   & 18.3   & 22.5   & 10.6   &  &  & 14.3     & 19.3     & 25.0     & 11.0    &  &  & 14.0     & 18.8     & 26.0     & 11.0    \\
   25000$^b$ &        &        &        &        &  &  &        &       &       &       &  &  &          &          &          &         \\
   27000 & 12.6   & 15.8   & 22.7   & 9.6    &    &  & 13.1     & 17.3     & 24.2     & 10.2    &  &  & 13.5     & 17.8     & 28.2     & 11.0    \\
   29000 & 15.8   & 22.6   & 27.3   & 12.2   &    &  & 16.1     & 23.1     & 28.2     & 12.5    &  &  & 19.6     & 25.8     & 30.4     & 15.0    \\
   31000 & 16.5   & 18.1   & 26.3   & 11.4   &  &  & 16.6     & 18.9     & 27.9     & 12.4    &  &  & 19.1     & 24.9     & 36.5     & 15.4    \\
   33000 & 19.7   & 24.9   & 25.5   & 14.2   &  &  & 21.0     & 26.1     & 29.0     & 15.5    &  &  & 24.3     & 31.1     & 49.2     & 20.5    \\ \hline
\end{tabular}
\ec
\tablecomments{0.86\textwidth}{$^a$The closest spectrum rather than itself is used as template.}
\tablecomments{0.86\textwidth}{$^b$The null values mean that there are no sample stars in these $T_{\mathrm{eff}}$ ranges.}
\end{table}


\section{Discussion}
\label{sect:D}

To sum up, the results show that the CSST spectroscopic survey is capable of delivering stellar radial velocities to a precision of $2 - 4 \,\mathrm{km}\,\mathrm{s}^{-1}$ for
AFGKM types of stars at SNR = 100. The precision decreases to $ 4 - 15 \,\mathrm{km}\,\mathrm{s}^{-1}$ for hotter OB stars.
The CSST spectroscopic survey aims to reach AB magnitude 5$\sigma$ limit  $\sim$21 mag per resolution element for point sources.
It corresponds to about 17 mag at SNR = 100. It means that the CSST spectroscopic survey has a potential to provide reliable velocities for
a huge number of stars for Galactic and stellar sciences, such as Galactic kinematics and  searching for high-velocity stars and hyper-velocity stars \citep{brown_hypervelocity_2015}.

There are several simplifications in the current work. We ignored any systematic uncertainties in the CSS-OS data reduction process (such as sky subtraction and CCD charge transfer inefficiency correction). 
We didn't consider variations of spectral resolution and SNRs as a function of wavelength either. 
We also ignored any uncertainties in the flux calibration processes. As mentioned in Section~\ref{sect:method}, it is highly possible to achieve a flux calibration precision of 1 per cent, 
using tens of millions of stars as flux standards.  Nevertheless, we can always use normalized spectra to derive stellar radial velocities, at a cost of slightly larger uncertainties.

Wavelength calibration plays an important role in the final precision of CSST radial velocities, in two ways. 
One is how precise are the field-dependent dispersion solutions, and another one is how precise are the wavelength zero points.
With a new star-based method that can monitor and correct for possible errors in the CSST dispersion solutions using normal scientific observations of stars of known velocities, 
Yuan et al. (2020) have shown that it is possible to achieve a precision of a few $\,\mathrm{km}\,\mathrm{s}^{-1}$ for the dispersion solutions. 
Here we discuss the precision of wavelength zero points $\delta v_{0}$ , which depends on  the precision of centroids of zero-order images $\Delta pixel$.

It is known that $\Delta pixel$ for a well-sampled stellar image can be written as $\Delta pixel = \frac{1.335\times FWHM}{SNR}$\citep{1978moas.coll..197L}, where $FWHM$ is in units of pixel.  
For the CSST spectrographs, the FWHM of their zero-order images is about 5.1 pixels (assuming a Gaussian profile) or smaller, given that the radius of 80\% energy concentration of zero-order images is no larger than 0.3 arc-second 
and the sampling is 0.074 arc-second per pixel. 
For a 17 mag star that has a typical spectral SNR of 100, it can be easily estimated that the SNRs of its zero-order images (dominated by Poisson noise) of 
GU/GV/GI bands are larger than 500, assuming a 5\% efficiency for the zero-order images and 20\%, 40\% and 40\% efficiency for the GU, GV and GI band spectra \citep{2018cosp...42E3821Z}.
Therefore, $\Delta pixel$ can be directly determined to a precision about 0.01 pixel, corresponding to a  $\delta v_{0}$ precision of about 4$\,\mathrm{km}\,\mathrm{s}^{-1}$.

Given the fact that almost all CSST stellar targets will have very accurate positions and proper motions thanks to the Gaia mission \citep{Lindegren_Gaia_2018A&A...616A...2L}, 
$\Delta pixel$ can also be estimated indirectly from a star's sky position. For a 17 mag star, Gaia DR5 will deliver positions accurate to about 0.05 mas 
and proper motion accurate to about 0.04 mas/yr \footnote{https://www.cosmos.esa.int/web/gaia/sp-figure2}. Assuming a typical epoch difference of 10 years between the Gaia DR5 and the CSST, 
the star will have a predicted position accurate to about 0.4 mas, corresponding to $\Delta pixel$ value of about 0.005 and a $\delta {v_{0}}$ precision of about 2$\,\mathrm{km}\,\mathrm{s}^{-1}$. Note that we ignore errors in the astrometric solutions,  which are 
supposed to be very small considering that a large number stars can be used. The above analyses show that possible velocity errors caused by wavelength zero points are comparable or smaller than 
 $\Delta v$ in our main results.  The conclusion also holds for fainter stars.

In addition, we assumed perfect template spectrum for a given target in our simulation. To investigate possible effect of imperfect template on velocity measurements, for each normalized spectrum, 
we have first tried every spectrum from the NGSL library rather than itself to find its closest spectrum as template to measure radial velocities, the results are shown in Figure 4 (lower-left panel), Figure 5 (right panel) 
and Figure 6 (lower-left panel). 
The uncertainties for most stars are almost the same to those using itself as template. The uncertainties are significantly larger for only few hot stars, due to lack of enough hot stars in the NGSL library. 
To further investigate additional uncertainties caused by imperfect template, we used a G2-type star ($\mathrm{HD 095241}$) as an example. We compared velocity uncertainties using itself as template with those using stars of different stellar parameters. 
Two stars ($\mathrm{BD+381670}$ and $\mathrm{HD 022484}$) were used. 
The $T_{\mathrm{eff}}$ differences are 243\,K and 94\,K, and the  
[Fe/H] differences are 0.23 and 0.24 dex, respectively. 
It's found that velocity uncertainties with imperfect template are only larger by about 10\%.  
The results suggest that the precision of velocities measurements is not sensitive to possible systematics in the templates.

The discussions above indicate that precision values obtained in the current work are upper limits. 
Many realistic situations have not been taken into account, such as lower SNRs, not cleanly background subtraction, 
overlapping (not only on spectra but also on zero-order images), no detected zero-order image, 
asymmetric and variable point spread function, charge transfer inefficiency etc.                                                                       
It is still unclear to what extent these complicated situations would affect the precision of the RV measured from slitless spectra.                
More investigations should be done with more dedicated simulated spectra considering all those situations in future works.

In the end, all targets in the NGSL are normal stars. However, the CSS-OS will observe various types of abnormal stars, such as emission line stars (young stellar objects, Ae stars, Be stars, cataclysmic variables) and white dwarfs. 
Due to the existence of strong emission lines or intrinsically very broad absorption features, the CSS-OS slitless spectra are very suitable to measure their velocities.

\section{Summary}
\label{sect:Sum}
By using the degraded NGSL spectra to stimulate the CSST stellar spectra, we have determined the best precision of  CSST radial velocities,
and its variations as functions of stellar parameters. We find that the best precision depends strongly on temperature, weakly on
metallicity for only solar-like stars, and hardly on surface gravity. At SNR=100, the best precision is about  $2 - 4 \,\mathrm{km}\,\mathrm{s}^{-1}$ for AFGKM  stars, and $4 -  15 \,\mathrm{km}\,\mathrm{s}^{-1}$ for OB stars.
The uncertainties of velocity measurement brought by wavelength calibration are considered, and are estimated to be around a few $\mathrm{km}\,\mathrm{s}^{-1}$.

Velocity uncertainties using single GU/GV/GI band spectra are also explored. At the same SNR, the GU band performs the best, the GV band the second best, and then the GI band. The effects of spectral normalization and imperfect template  on velocity measurements are investigated, and found to be  weak if done properly. Given the precision of radial velocities and the large number of stars that can be observed, the CSST spectroscopic survey can enable interesting science such as searching for high-velocity stars and hyper-velocity stars.

 We have made several simplifications that the results in the work are the upper limits of the real cases. To make the numbers come true, lots of efforts are required to achieve accurate 
 wavelength calibration, particularly in the GU band.
In future we will re-visit the precision of velocity measurements using more realistic simulated CSS-OS data when available and extend to other stellar parameters, 
including temperature, metallicity, surface gravity, alpha  and other elemental abundances.

\normalem
\begin{acknowledgements}
 Dedicated to the Department of Astronomy of Beijing Normal University, the 2nd astronomy program in the modern history of China. 
 We acknowledge the referee for his/her valuable comments that improved the quality of this work significantly.
 This work is supported by the National Key Basic R\&D Program of China via 2019YFA0405500, the National Natural Science Foundation of China through the project NSFC 11603002 and 
 Beijing Normal University grant No. 310232102.
\end{acknowledgements}

\bibliographystyle{raa}
\bibliography{2020-0278}

\begin{thebibliography}{15}
\providecommand\natexlab[1]{#1}
\providecommand\JournalTitle[1]{#1}

\bibitem[{Baranne} {et~al.}(1996)]{1996A&AS..119..373B}
{Baranne}, A., {Queloz}, D., {Mayor}, M., {et~al.} 1996, \aaps, 119, 373

\bibitem[Brown(2015)]{brown_hypervelocity_2015}
Brown, W.~R. 2015, Annual Review of Astronomy and Astrophysics, 53, 15

\bibitem[Cao {et~al.}(2018)]{cao_testing_2018}
Cao, Y., Gong, Y., Meng, X.-M., {et~al.} 2018, Monthly Notices of the Royal
  Astronomical Society

\bibitem[Glazebrook {et~al.}(2005)]{glazebrook_monster_2005}
Glazebrook, K., Baldry, I., Moos, W., Kruk, J., \& McCandliss, S. 2005, New
  Astronomy Reviews, 49, 374, arXiv: astro-ph/0410037

\bibitem[Gong {et~al.}(2019)]{gong_cosmology_2019}
Gong, Y., Liu, X., Cao, Y., {et~al.} 2019, The Astrophysical Journal, 883, 203,
  arXiv: 1901.04634

\bibitem[{Heap} \& {Lindler}(2007)]{2007ASPC..374..409H}
{Heap}, S.~R., \& {Lindler}, D.~J. 2007, Astronomical Society of the Pacific
  Conference Series, Vol. 374, {Hubble's Next Generation Spectral Library
  (NGSL)}, ed. A.~{Vallenari}, R.~{Tantalo}, L.~{Portinari}, \& A.~{Moretti},
  Astronomical Society of the Pacific Conference Series, Vol. 374, From Stars
  to Galaxies: Building the Pieces to Build Up the Universe, ed.
  A.~{Vallenari}, R.~{Tantalo}, L.~{Portinari}, \& A.~{Moretti}, 409

\bibitem[Irfan {et~al.}(2019)]{irfan_determining_2019}
Irfan, M.~O., Bobin, J., Miville-Deschênes, M.-A., \& Grenier, I. 2019,
  Astronomy \& Astrophysics, 623, A21

\bibitem[Koleva \& Vazdekis(2012)]{koleva_stellar_2012}
Koleva, M., \& Vazdekis, A.~V. 2012, Astronomy \& Astrophysics, 538, A143,
  arXiv: 1111.5449

\bibitem[{Lindegren}(1978)]{1978moas.coll..197L}
{Lindegren}, L. 1978, in IAU Colloq. 48: Modern Astrometry, 197

\bibitem[{Lindegren} {et~al.}(2018)]{Lindegren_Gaia_2018A&A...616A...2L}
{Lindegren}, L., {Hern{\'a}ndez}, J., {Bombrun}, A., {et~al.} 2018, \aap, 616,
  A2

\bibitem[{Planck Collaboration}
  {et~al.}(2014)]{planck_collaboration_planck_2014}
{Planck Collaboration}, Abergel, A., Ade, P. A.~R., {et~al.} 2014, Astronomy \&
  Astrophysics, 571, A11

\bibitem[Schlegel {et~al.}(1998)]{schlegel_application_1998}
Schlegel, D., Finkbeiner, D., \& Davis, M. 1998, Wide Field Surveys in
  Cosmology, 14, 297

\bibitem[Ting {et~al.}(2017)]{ting_prospects_2017}
Ting, Y.-S., Conroy, C., Rix, H.-W., \& Cargile, P. 2017, The Astrophysical
  Journal, 843, 32

\bibitem[Zhan(2011)]{zhan_consideration_2011}
Zhan, H. 2011, SCIENTIA SINICA Physica, Mechanica \& Astronomica, 41, 1441

\bibitem[{Zhan}(2018)]{2018cosp...42E3821Z}
{Zhan}, H. 2018, in 42nd COSPAR Scientific Assembly, Vol.~42, E1.16

\end{thebibliography}


\end{document}